\documentstyle[twoside,fleqn,espcrc2,axodraw]{article}

\def\a{\alpha}

\def\e{\epsilon}
\def\f{\phi}

\def\m{\mu}
\def\n{\nu}

\def\s{\sigma}

\def\G{\Gamma}

\def\co{{\cal O}}

\def\beql{\begin{eqnarray}}
\def\eeql{\end{eqnarray}}
\def\NO{\nonumber}

\def\sign(#1){(\!-\!1)^{#1}}
\def\binom(#1,#2){ (\!\!
	 \begin{array}{c} #1 \\ #2 \end{array}\!\! ) }
\def\plus{\!+\!}
\def\minus{\!-\!}
\def\mydot{\!\!\cdot\!}
\def\nn{\nonumber \\ &&}

\def\TAC(#1,#2,#3,#4,#5,#6){
        \raisebox{-19.1pt}{
        \SetScale{0.5} \SetPFont{Helvetica}{14}
        \hspace{-15pt}
        \begin{picture}(50,39)(0,-4)
        \SetColor{Blue}
        \CArc(40,35)(25,90,270) \CArc(60,35)(25,270,90)
        \Line(40,60)(60,60) \Line(40,10)(60,10) \Line(50,10)(50,60)
        \Line(0,35)(15,35) \Line(85,35)(100,35)
        \SetColor{Black}
        \SetColor{Red}
        \SetWidth{3}
                \Line(40,60)(50,60)
                \CArc(40,35)(25,90,180)
        \SetWidth{0.5}
        \end{picture}
        \SetScale{1.0}
        \hspace{-7pt}
        }
}
\def\TAD(#1,#2,#3,#4,#5,#6){
        \raisebox{-19.1pt}{
        \SetScale{0.5} \SetPFont{Helvetica}{14}
        \hspace{-15pt}
        \begin{picture}(50,39)(0,-4)
        \SetColor{Blue}
        \CArc(40,35)(25,90,270) \CArc(60,35)(25,270,90)
        \Line(40,60)(60,60) \Line(40,10)(60,10) \Line(50,10)(50,60)
        \Line(0,35)(15,35) \Line(85,35)(100,35)
        \SetColor{Black}
        \SetColor{Red}
        \SetWidth{3}
                \Line(50,10)(50,60)
        \SetWidth{0.5}
        \end{picture}
        \SetScale{1.0}
        \hspace{-7pt}
        }
}
\def\TAF(#1,#2,#3,#4,#5,#6,#7){
	\raisebox{-19.1pt}{
	\SetScale{0.5} \SetPFont{Helvetica}{14}
	\hspace{-15pt}
	\begin{picture}(50,39)(0,-4)
	\SetColor{Blue}
	\CArc(40,35)(25,90,270) \CArc(60,35)(25,270,90)
	\Line(40,60)(60,60) \Line(40,10)(60,10) \Line(50,10)(50,60)
	\Line(0,35)(15,35) \Line(85,35)(100,35)
	\SetColor{Black}
	\PText(53,40)(0)[l]{#6}
	\PText(40,65)(0)[rb]{#1} \PText(15,48)(0)[rb]{#7} 
	\PText(65,62)(0)[lb]{#2}
	\PText(65,12)(0)[lt]{#3} \PText(35,12)(0)[rt]{#4}
	\SetColor{Red}
	\SetWidth{3}
		\Line(50,10)(50,60)
		\Line(40,60)(50,60)
		\CArc(40,35)(25,90,130)
		\Vertex(50,60){1.3}
	\SetWidth{0.5}
	\end{picture}
	\SetScale{1.0}
	\hspace{-7pt}
	}
}
\def\TAL(#1,#2,#3,#4,#5,#6,#7){
	\raisebox{-19.1pt}{
	\SetScale{0.5} \SetPFont{Helvetica}{14}
	\hspace{-15pt}
	\begin{picture}(50,39)(0,-4)
	\SetColor{Blue}
	\CArc(40,35)(25,90,270) \CArc(60,35)(25,270,90)
	\Line(40,60)(60,60) \Line(40,10)(60,10) \Line(50,10)(50,60)
	\Line(0,35)(15,35) \Line(85,35)(100,35)
	\SetColor{Black}
	\PText(53,40)(0)[l]{#5}
	\PText(40,65)(0)[rb]{#1} \PText(15,48)(0)[rb]{#6}
	\PText(65,62)(0)[lb]{#2}
	\PText(60,9)(0)[lt]{#3} \PText(85,26)(0)[lt]{#7}
	\PText(35,12)(0)[rt]{#4}
	\SetColor{Red}
	\SetWidth{3}
		\Line(50,10)(50,60)
		\Line(40,60)(50,60)
		\CArc(40,35)(25,90,130)
		\Vertex(50,60){1.3}
		\Line(50,10)(60,10)
		\CArc(60,35)(25,270,310)
		\Vertex(50,10){1.3}
	\SetWidth{0.5}
	\end{picture}
	\SetScale{1.0}
	\hspace{-7pt}
	}
}

\title{
\vspace*{-35mm}
\rightline{
{\tenrm{NIKHEF 99-022}}}
\vspace*{-2mm}
\rightline{\tenrm{September 1999}}
\vspace*{+20mm}
The Mellin moments of deep inelastic structure functions 
	at two loops}

\author{S. Moch{\thanks{Talk at the QCD 99 Euroconference,  
	7-13th July 1999, Montpellier (France).}} 
	and J.A.M. Vermaseren\\
	NIKHEF Theory Group, 
	Kruislaan 409, 1098 SJ Amsterdam, The Netherlands}

\begin{document}
\pagestyle{empty}

\begin{abstract}
We perform the analytic calculation of the Mellin moments 
of the structure functions $F_2$ and $F_L$ in perturbative 
QCD up to second order corrections and in leading twist 
approximation. We calculate the 2-loop contributions to the 
anomalous dimensions of the singlet and non-singlet 
operator matrix elements and the 2-loop coefficient functions 
of $F_2$ and $F_L$. Our results are in agreement with earlier 
calculations in the literature.
\end{abstract}

\maketitle

\section{INTRODUCTION}

Deep inelastic lepton-hadron scattering is one of the best studied 
reactions today. It provides unique 
information about the structure of the hadrons and tests one of the 
most important predictions of perturbative QCD, the scale evolution 
of the structure functions \cite{gw73,bbdm78,gpp1}. 
At the same time, the ever increasing accuracy of 
deep inelastic scattering (DIS) experiments demands more 
accurate theoretical predictions and despite all achievements 
\cite{kk88,cLpp2,lv93,zvn92}, no complete next-to-next-to-leading order 
(NNLO) analysis for DIS reactions is available. 

The necessary perturbative QCD predictions at 3-loops for the anomalous dimensions 
of the structure functions and the structure function $F_L$ entering 
in the ratio $R = \s_L / \s_T$, are still unknown, except for a 
number of fixed Mellin moments of $F_2$ and $F_L$ \cite{lrv94,lnrv97}.  
These have already been used in NNLO analyses \cite{lnrv97,kkps98,sy99}.
In general however, this limited information about some fixed Mellin moments 
is not sufficient to allow for NNLO analyses of all data of DIS experiments 
and as a consequence, there are still considerable uncertainties 
on the parton densities, e.g. on the gluon density at small $x$. 

Progress beyond the state of the art has to 
explore different directions, since a straightforward extension of ref.\cite{lnrv97} 
to calculate more fixed Mellin moments, is not feasible. 
One such possibility that actually dates back to the origins 
of QCD \cite{gw73,gpp1}, is to calculate the Mellin moments 
of the structure functions analytically as a general function of $N$. 
This approach, further pioneered by Kazakov and Kotikov \cite{kk88} 
to obtain the structure function $F_L$ at 2-loops, turns out to be 
flexible and very promising in view of the 
ultimate goal, the anomalous dimensions at 3-loops. 
As a first step and to demonstrate the power of the method, 
we report here on the recalculation \cite{mvtbp} of the perturbative QCD 
contributions up to 2-loops to the structure functions 
$F_2$ and $F_L$.

\section{FORMALISM}

The optical theorem relates the DIS structure functions to the 
forward scattering amplitude of photon-nucleon scattering, $T_{\m\n}$, 
which has a time-ordered product of two local electromagnetic currents, 
$j_\m(x)$ and $j_\n(z)$, Fourier transformed into momentum space, 
to which standard perturbation theory applies. 
The operator product expansion (OPE) allows 
to expand this current product around the lightcone $(x-z)^2 \sim 0$ 
into a series of local composite operators of leading twist and spin $N$. 
The anomalous dimensions of matrix elements of these operators 
govern the scale evolution of the structure functions, while 
the coefficient functions multiplying these matrix elements 
are calculable order by order in perturbative QCD.

In this way the Mellin moments of DIS structure functions can 
naturally be written in the parameters of the OPE,  
\beql
\label{eq:F2mellin}
{\lefteqn{
\frac{1+(-1)^N}{2} \int\limits_0^1 dx\, x^{N-2} F_i(x,Q^2)
= }} \\
& &\sum\limits_{k={\rm{ns, q, g}}}
C_{i,N}^k\left(Q^2/\m^2,\a_s\right) A_{{\rm{nucl}},N}^k(\m^2)\, , 
\NO
\eeql
where $i=2,L$ and all even Mellin moments are fixed. 
$A_{{\rm{nucl}}}$ denote the spin-averaged hadronic 
matrix elements and $C_{i}$ the coefficient functions and 
the sum extends over the flavour non-singlet and 
singlet quark and gluon contributions. 

The OPE is an operator statement and therefore independent 
of a particular hadronic matrix element, so that it is standard 
to calculate partonic structure functions 
with external quarks and gluons in infrared regulated perturbation theory. 
In practice, this procedure reduces to the task of calculating the 
$N$-th moment of all 4-point diagrams that contribute to $T_{\m\n}$ 
at a given order in perturbation theory. 
This will be shown in more detail below.

\section{METHOD}

To illustrate the method of calculating Mellin moments 
of DIS structure functions, 
consider the following diagram, which is the most complex 
topology at 2-loops, since three propagators with $P$-momentum 
are present. 
\beql
\label{eq:TAL}
{\lefteqn{
        \TAL({A},{b},{C},{d},{E},{a},{c}) = 
\int\!\! d^Dp_1 d^Dp_2 
		\frac{1}{(p_1^2)^a((P\plus p_1)^2)^A}
}}\\
& &\times\,
                \frac{1}{(p_2^2)^b(p_3^2)^c
		((P\minus p_3)^2)^C(p_4^2)^d((P\plus p_5)^2)^E}\, .
\NO
\eeql
The fat lines denote $P$-dependent propagators with $P^2 = 0$, 
while the $Q$-momentum flows from right to left through the diagram.  
Usually, there are only single powers of the propagators, 
$a=b= ... =1$ in eq.(\ref{eq:TAL}).

We wish to compute the $N$-th moment of this diagram. 
In a naive approach one would expand all 
$P$-dependent propagators into sums over $P\mydot p_i/p_i^2$ 
using $P^2 = 0$. Then scaling arguments require the final 
answer for the $N$-th moment to be proportional 
to $(P\mydot Q/Q^2)^N$ 
and one is left to calculate 2-point functions with symbolic 
powers of scalar products in the numerator and denominator. 
However, this procedure leads to multiple nested sums, which 
in general are very difficult to evaluate. 

More sophisticated ideas have been used in ref.\cite{kk88}. 
There it is shown how to determine reduction identities 
for a given diagram in dimensional regularization in such a way, 
that is possible to set up recursion relations in the number of moments $N$.
A systematic classification of all 2-loop topologies 
reveals two basic building blocks. 
The two diagrams, each with only one $P$-dependent propagator,
\beql
\label{eq:TAC/TAD}
	&\quad &
	\TAC({A},{b},{C},{d},{E},{a},{c})\, , \quad\quad
	\TAD({A},{b},{C},{d},{E},{a},{c})\, ,
\eeql
are shown in diagrammatic notation similar to eq.(\ref{eq:TAL}). 
These diagrams can be calculated at the cost of one sum 
over $\G$-functions. 
In dimensional regularization, $D=4-2\e$, the $\G$-functions 
can be expanded in $\e$ and the sum can be solved to any order in $\e$.

The other more complex topologies like eq.(\ref{eq:TAL}) 
can be expressed in sums over these basic topologies 
by means of recursions, so that a complex diagram 
is broken up systematically into its building blocks.
For example, the simple and beautiful recursion relation 
for eq.(\ref{eq:TAL}) reads
\beql
\label{eq:niceexample}
{\lefteqn{
	\TAL({1},{1},{1},{1},{1},{1},{1})(\frac{2 P\mydot Q}{Q\mydot Q})^k  \,=}}\\
& &
		-\frac{N\minus k\minus D\plus 6}{N\minus k\plus 2}
				\TAL(1,1,1,1,1,1,1)(\frac{2 P\mydot Q}{Q\mydot Q})^{k\plus 1}
		\nn
	+\frac{2}{N\minus k\plus 2}\TAF(1,1,2,1,0,1,1)(\frac{2 P\mydot Q}{Q\mydot Q})^k\, ,
\nonumber
\eeql
which relates the $N$-th moment of this diagram to a diagram of a simpler 
structure, where the $P$-dependent propagator $(P\minus p_3)^2$ has been 
eliminated. 
The recursion eq.(\ref{eq:niceexample}) can be solved and for the second 
diagram on the right hand side similar relations are derived.

Amazingly enough, if the recursion relations are set up in this way, 
all the multiple nested sums can all be solved successively 
in terms of harmonic sums, 
the basic functions of weight $m$ being defined as follows \cite{summer,b99}
\beql
\label{eq:basicharmo}
S_m(N) = \sum\limits_{i=1}^N  \frac{1}{i^m}\, , \quad 
S_{-m}(N) = \sum\limits_{i=1}^N  \frac{(-1)^m}{i^m}\, ,
\eeql
while higher functions can be defined recursively
\beql
\label{eq:higherharmo1}
S_{m_1,...,m_k}(N) \!&=&\! 
	\sum\limits_{i=1}^N  \frac{1}{i^{m_1}} S_{m_2,...,m_k}(i)\, , \\
\label{eq:higherharmo2}
S_{-m_1,...,m_k}(N) \!&=&\! 
	\sum\limits_{i=1}^N  \frac{(-1)^{m_1}}{i^{m_1}} S_{m_2,...,m_k}(i)\, .
\eeql

Of course, this procedure requires great care in the way the reduction 
identies are applied. 
It also relies crucially on all algebraic relations 
for harmonic sums, which allow to solve the nested sums algorithmically 
\cite{summer} and to express the 
result in the basis of harmonic sums 
eqs.(\ref{eq:basicharmo})--(\ref{eq:higherharmo2}).

\section{RESULTS}

The analytic calculation has been done with the symbolic manipulation 
program FORM \cite {form}. 
All recursion relations have been implemented in a program, 
that reduces diagrams for DIS structure functions up to 2-loops 
to multiple nested sums over the basic building blocks 
and subsequently calls the SUMMER \cite{summer} algorithm, 
to solve these nested sums in terms of the basis of harmonic sums. 
The database of diagrams, identical to the one used in ref.\cite{lnrv97},
was generated with the help of QGRAF \cite{qgraf}.
The program has been optimized by tabulating basic building blocks, 
so that the calculation of all 425 diagrams for $F_2$ and $F_L$ 
up to 2-loops is done in a few of hours on a Pentium Pro PC.

The calculation is performed in dimensional regularization, $D=4-2\e$, and 
renormalization and mass factorization proceeds as described 
in refs.\cite{lv93,zvn92,lrv94,lnrv97}. 
To extract the gluon anomalous dimension at 2-loops, we also calculate 
the unphysical structure functions of a scalar particle $\f$, 
that couples to gluons only \cite{lnrv97}.
These provide us with the necessary renormalization constants 
of the singlet operators in eq.(\ref{eq:F2mellin}) and 
can be obtained from an additional operator 
$\f F^{a\,\m\n} F^a_{\m\n}$ in the Lagrangian.

Finally, we obtain the even Mellin moments of the 
complete set of anomalous dimensions, the flavour non-singlet and singlet quark 
and the gluon coefficent functions up to 2-loops. 
Our results for the even moments of the anomalous dimensions 
agree with the ones published in refs.\cite{gw73,gpp1}, while  
our results for the coefficient functions agree with 
those of refs.\cite{bbdm78,lv93,zvn92,lrv94,lnrv97}. 

We have also calculated all even moments of the $\co(\e)$-terms 
of all structure functions at 2-loops, needed to extract 
the coefficient functions at 3-loops after mass factorization. 
In the appendix, we present our result for the even moments of the 
pure singlet quark coefficent function $c_2^{\rm{ps}}$, 
which has not been shown in this form before. 
The complete results and more details of the calculation 
will be published elsewhere \cite{mvtbp}.

\subsection*{Acknowledgments}
This work is part of the research program of the
Foundation for Fundamental Research of Matter (FOM) and
the National Organization for Scientific Research (NWO).   

\subsection*{APPENDIX}
The even Mellin moments of the pure singlet quark coefficient 
function $c_2^{\rm{ps}}$ at 2-loops.

\beql
{\lefteqn{
c_2^{\rm{ps}} = n_f C_F \bigl[ }} \\
& &
 - 133/81\ \delta(N\minus2)\ 
+ \theta(N\minus3)\ \bigl\{
\nn
          - 344/27 S_{1}(N\minus 2)
          - 130/27 S_{1}(N\minus 1)
\nn
          - 1714/27 S_{1}(N\plus 1)
          + 448/27 S_{1}(N\plus 2)
\nn
          + 580/9 S_{1}(N)
          - 16/3 S_{1,-2}(N\minus 2)
\nn
          + 64/3 S_{1,-2}(N\minus 1)
          + 64/3 S_{1,-2}(N\plus 1)
\nn
          - 16/3 S_{1,-2}(N\plus 2)
          - 32 S_{1,-2}(N)
\nn
          + 104/9 S_{1,1}(N\minus 2)
          - 416/9 S_{1,1}(N\minus 1)
\nn
          - 272/9 S_{1,1}(N\plus 1)
          + 32/9 S_{1,1}(N\plus 2)
\nn
          + 184/3 S_{1,1}(N)
          - 16/3 S_{1,1,1}(N\minus 2)
\nn
          + 4/3 S_{1,1,1}(N\minus 1)
          + 4/3 S_{1,1,1}(N\plus 1)
\nn
          - 16/3 S_{1,1,1}(N\plus 2)
          + 8 S_{1,1,1}(N)
\nn
          + 16/3 S_{1,2}(N\minus 2)
          - 4/3 S_{1,2}(N\minus 1)
\nn
          - 4/3 S_{1,2}(N\plus 1)
          + 16/3 S_{1,2}(N\plus 2)
\nn
          - 8 S_{1,2}(N)
          + 56 S_{2}(N\minus 1)
\nn
          + 136/9 S_{2}(N\plus 1)
          + 128/9 S_{2}(N\plus 2)
\nn
          - 256/3 S_{2}(N)
          - 16 S_{2,1}(N\plus 1)
\nn
          + 16 S_{2,1}(N\plus 2)
          + 8 S_{2,1,1}(N\minus 1)
\nn
          - 8 S_{2,1,1}(N\plus 1)
          - 8 S_{2,2}(N\minus 1)
\nn
          + 8 S_{2,2}(N\plus 1)
          + 2 S_{3}(N\minus 1)
\nn
          + 154/3 S_{3}(N\plus 1)
          - 64/3 S_{3}(N\plus 2)
\nn
          - 32 S_{3}(N)
          - 16 S_{3,1}(N\minus 1)
\nn
          + 16 S_{3,1}(N\plus 1)
          + 20 S_{4}(N\minus 1)
\nn
          - 20 S_{4}(N\plus 1)
	\bigr\} \bigr]\, . \NO
\eeql

\end{document}